# Evaluation of Decoherence for Quantum Computing Architectures: Qubit System Subject to Time-Dependent Control

Dmitry Solenov and Vladimir Privman

Center for Quantum Device Technology,
Department of Physics, Clarkson University, Potsdam, New York 13729–5721, USA
Electronic addresses: solenov@clarkson.edu, privman@clarkson.edu

We present an approach that allows quantifying decoherence processes in an open quantum system subject to external time-dependent control. Interactions with the environment are modeled by a standard bosonic heat bath. We develop two unitarity-preserving approximation schemes to calculate the reduced density matrix. One of the approximations relies on a short-time factorization of the evolution operator, while the other utilizes expansion in terms of the system-bath coupling strength. Applications are reported for two illustrative systems: an exactly solvable adiabatic model, and a model of a rotating-wave quantum-computing gate function. The approximations are found to produce consistent results at short and intermediate times.

## 1. Introduction

A wide range of realizations of qubit designs have been proposed[1-10] for solid-state quantum computing. Subject to inevitable interaction with the environment, a qubit system deviates from coherent evolution, developing entanglement with the environmental modes. The goal of quantum information processing is to manipulate a set of qubits keeping a relatively low degree of the resulting noise-induced error. A variety of techniques have been proposed to deal with this issue, including quantum error correction codes,[11-15] decoherence-free subspaces,[16-18] and dynamical decoupling.[19,20] In parallel, approaches to quantifying deviations from coherent evolution have been developed,[21-45] providing tools for investigating the environmentally induced noise in quantum computing architectures. In most cases, it is impossible to obtain the exact evolution for the reduced density matrix of the qubits with the environmental modes traced over. Recent studies of decoherence have therefore focused on different simplified models and approximations that allow to quantify decoherence processes on various time scales[28,29,32,39,42] in order to evaluate measures to decoherence to compare with the noise threshold requirements of quantum error correction schemes.

Generally, quantum computing algorithms rely on the possibility to apply a certain number of gates consecutively on a qubit register.[46] The details of the interactions used for each gate are irrelevant to the actual computation as long as the final evolution operator of the gate has the desired form.[46,47] However, in any practical realization in a physical system of qubits (two-state quantum systems), a number of factors may affect the actual choice of pulse shapes used for quantum gates.[48]

Most theoretical studies of decoherence have been restricted to idling qubits, because time-independent interactions are easier to handle. Estimates of the noise level for such systems have been assumed to give an indication of the behavior for more complicated gated-qubit situations. Attempts to handle time-dependent interactions were primarily made[49,50] for Markovian-type models,[51-53] which are applicable for relatively large times for which the environmental modes can be viewed as a thermalized system. However, for fast quantum computing dynamics, at short to intermediate time scales, different models are more appropriate.[29]

In this work we introduce two approaches to evaluation of decoherence for qubits subject to fast time-dependent gate operations on time scales appropriate for quantum computing. The standard paradigm for quantum computation involves a succession of single- and two-qubit gates. The specific results reported in this work are for a single qubit interacting with environment. The developed formalism can be generalized to handle several-qubit gates as well. However, the calculations will be prohibitively complicated. Instead, approaches relying on approximate additivity[31,32,36] of noise measures could be more appropriate.

The Hamiltonian of the system is
$$H = H_S(t) + H_B + H_{SB} ,\tag{1.1}$$
where
$$H_S(t) = H_i + H_g(t) .\tag{1.2}$$
Here $H_i$ is the Hamiltonian of an idling qubit, while $H_g(t)$ represents time-dependent gate control. The Hamiltonian of the environment is given by $H_B$, and $H_{SB}$ is its interaction with the qubit. The environment is often modeled as a bath of non-interacting modes,[54-60] with Hamiltonians $M_k$,
$$H_B = \sum_k M_k .\tag{1.3}$$
The system-bath interaction term is usually modeled [60] as
$$H_{SB} = S \sum_k X_k ,\tag{1.4}$$
where $S$ is a Hermitian operator in the system space, coupled to the bath operators $X_k$. We will focus on the bosonic bath,[54-60]
$$M_k = \omega_k a_k^\dagger a_k ,\tag{1.5}$$
where we set $\hbar = 1$ to simplify the notation. The standard choice of the linear coupling[60] is
$$X_k = g_k a_k^\dagger + g_k^* a_k .\tag{1.6}$$

Systems candidate for quantum computing are usually well isolated from the environment, and it is often expected that the interaction term is small with respect to other energy scales of the problem.[14,15,39,42,51-53] Therefore, many approaches to evaluating decoherence utilize various forms of perturbative approximation in powers of $|g_k|$. Furthermore, approximation techniques appropriate for short times, $t$, with the time variable itself defining "small" quantities once the fast dynamics has been factored out, have also been developed[29] and found many applications.[61-63] These two types of approach—perturbative in time or in interaction—are the only presently available treatments of decoherence at short times. Here we will analyze and compare both approaches, developing the technique to accommodate time-dependent external "gate" interaction.

Quantum computing involves a sequence of time intervals during which operations are performed on single qubits or on pairs of qubits.[46] For practical reasons of tractability it is usually assumed[26,39,40,42,60] that evaluation of the degree of decoherence for a given time interval can be approximately carried out assuming that initially the qubit system is not entangled with the bath, i.e., the overall density matrix, $\rho(t)$, is factorized at $t = 0$,

$$\rho(0) = \rho_S(0)\rho_B(0) . \tag{1.7}$$

Furthermore, it is usually assumed[60] that initially the bath is in thermal equilibrium,

$$\rho_B(0) = e^{-H_B/kT} \Big/ Tr_B\left(e^{-H_B/kT}\right) . \tag{1.8}$$

The reduced density matrix, $\rho_S(t) = Tr_B[\rho(t)]$, then gives the evolution of the qubit system at times $t \geq 0$.

The approach to evaluation of decoherence with the environment modeled as a closed (except for the interaction with the qubit) system of noninteracting modes, is often called a "short-time" treatment of decoherence. It has been utilized[29] as a reasonable approximation at low temperatures, at which most of the bath relaxation time scales are large as compared to fast qubit manipulations. At longer times, however, the bath mode interactions with a larger surrounding, as well as the interaction between the bath modes themselves, will result in the destruction of the bath's fragile entanglement with the qubit. The latter behavior has been modeled[51-53] by resetting the bath modes to the thermalized state (1.8). This "Markovian approach" to decoherence provides a treatment appropriate for larger times, when one assumes that the bath modes are thus reset to thermal state after each infinitesimal time step. Specifically, it has been argued[23,33,35] that for "pure" decoherence, dominated by processes that do not involve energy exchange with the bath, the crossover between the two regimes occurs on time scales of order $1/kT$ $(= \hbar/kT)$. Here we work in the "short-time" regime.

Introduction of the time-dependent gate-function interaction, $H_g(t)$, see (1.2), makes the short-time techniques for evaluating decoherence developed in Ref. 29 inapplicable. In this work we developed a short-time approximation formalism applicable for time-dependent interactions. As in the time-independent short-time approach, our new formulation preserves unitarity of the qubit-bath evolution and reproduces exact results for a solvable adiabatic-coupling model. We also present an alternative approach based on a unitarity-preserving perturbative expansion in terms of the qubit-bath coupling strength.

In Section 2, we derive our new short-time approximation scheme and discuss its limits of validity. An explicit expression for the reduced density matrix is given. In Section 3, we set up the formalism for the second-order perturbative approximation scheme for calculating the evolution operator of the qubit-bath system utilizing the Magnus expansion,[64,65] which allows us to avoid dealing with time-ordering. Expression for the reduced density matrix of a qubit is obtained, and the appropriate traces over the bath modes required for its evaluation are calculated.

In Section 4, the developed approximations are applied to the adiabatic model.[26] In this case our short-time scheme reproduces the exact solution while the perturbative approach follows the exact result up to the expected, order $|g|^3$, corrections. Our main result is presented in Section 5, in which we evaluate decoherence of a qubit controlled by an external rotating wave.[66,67] The model of rotating wave control provides a relatively simple, though quite general example of dealing with decoherence in the presence of a time-dependent gate function. Nonlinear dependence of decoherence properties on the rotating wave amplitude is analyzed.

## 2. Short-time approximation for time-dependent gate interactions

The recipe for constructing the reduced density matrix at short times has been developed for "idling" qubit systems described by time-independent Hamiltonians.[29] It is based on an approximate factorization of the overall evolution operator which allows tracing over the bath variables. The effect of the bath is then lumped into complex weight factors[26,29] entering the expression for the reduced density matrix, which completely describe the short-time decoherence processes. In this section we develop a formulation of the short-time approximation appropriate for analyzing decoherence for the time-dependent case.

We begin with the assumption that the evolution of the coherent system, with the Hamiltonian $H_S(t)$, is known. Specifically, for a single-qubit (two-state) system one can obtain, even with a time-dependent Hamiltonian, either numerically or, in some cases, analytically, accurate expressions[68-72] for the evolution operator, $U_S(t)$. Generally, we will assume that the solution of

$$i\dot{U}_S(t) = H_S(t)U_S(t) \tag{2.1}$$

is available. The overall density matrix for $t > 0$ can be obtained in terms of the evolution operator, $U_T(t)$,

$$\rho(t) = U_T(t)\rho(0)U_T^\dagger(t) \ . \tag{2.2}$$

The latter is the solution of

$$i\dot{U}_T(t) = [H_S(t) + H_B + H_{SB}]U_T(t) \ . \tag{2.3}$$

For short times, we approximately factor the evolution operator $U_T(t)$ into a product that allows to compute the trace over the bath modes of the expression (2.2), yielding the reduced density matrix. We propose the ansatz

$$U_T(t) = W_S(t)V(t)W_S(t) \ , \tag{2.4}$$

where the unitary operator $W_S(t) = \sqrt{U_S(t)}$ was introduced. Square root of a unitary operator is generally not a uniquely defined object due to the possibility to include phases in multiplies of $\pi$, in the diagonalised representation. This ambiguity, however, does not affect further derivations. Substituting (2.4) in (2.3) yields the equation for $V(t)$,

$$i\dot{V}(t) = (H_B + H_{SB})V(t) + W_S^\dagger(t)[H_{SB}, W_S(t)]V(t) + [i\dot{W}_S(t)W_S^\dagger(t), V(t)] \ . \quad (2.5)$$

When dealing with idling qubits, i.e., no external gate is applied, there are two physical processes that contribute to decoherence.[26,29,51,52] One of them involves qubit-bath energy exchange and leads to redistribution of the probabilities for the system to be in its various energy states, as well as to decoherence. For large times, such "resonant" relaxation, when described within an appropriately formulated Markovian approximation for the bath modes, yields a fully diagonal, in the energy basis, thermalized reduced density matrix. The other process, often called "pure decoherence," does not involve transfer of energy but results in phase noise that disrupts coherent evolution of the system. Since such nonresonant decoherence does not involve any specific time scales, it can be dominant in disrupting quantum computation at short times. Furthermore, as described in the Introduction, it has been argued[23,33,35] that the time scale of the onset of Markovian re-thermalization of the bath modes is given by $\hbar/kT$. This time can be quite large for low-temperature systems utilized in quantum computing applications. Therefore, evaluation of short-time decoherence usually involves no Markovian assumption. In addition, an approximation is invoked of considering only pure decoherence,[26,29] or some other perturbative approach is used,[39,42] in order to make calculations tractable.

The former, pure-decoherence approximation[29] for the time-independent case, $H_S(t) = const$, corresponds to the assumption that $H_S$ is approximately conserved, $[H_S, H] \simeq 0$, namely that the system operators $H_S$ and $S$ approximately commute with each other. Therefore, one can omit the second and the third terms in (2.5). This effectively corresponds to a short-time approximation that retains terms of order $t^2$ and preserves the unitarity of the dynamics. We get

$$V(t) = e^{-i(H_B + H_{SB})t} \ , \quad (2.6)$$

which reproduces the short-time approximation of Ref. 29, derived by a different method. Note that even though expression (2.6) has terms with all the powers of time, only those up to $t^2$ inclusive, coincide with the exact solution, while the higher powers are kept to preserve the unitarity of the dynamics. In this section, we consider pure-decoherence type approximations, while Section 3 is devoted to a different, perturbative, approximation. When $H_S(t)$ is time-dependent, generally the system's energy levels as well as its energy exchange with the bath, and, therefore, separation of resonant vs. pure-decoherence relaxation processes, are not well defined. Nevertheless, we will demonstrate that in most situations omitting the last two terms in (2.5) still provides a unitarity-preserving approximation retaining terms of order $t^2$, inclusive.

Let us tentatively assume that the last two terms in (2.5) are small and write

$$V(t) = e^{-i(H_B + H_{SB})t}[1 + R(t)] \ . \quad (2.7)$$

By an iterative expansion of (2.5), one can show that the leading-order corrections to the approximate expression (2.6) have the form

$$R(t) = \frac{i}{2}[\int_0^t dt' \int_0^{t'} dt'' H_g(t'') - \int_0^t dt' H_g(t')t', H_{SB}] + O(\varepsilon^2 |g| t^3) \ , \quad (2.8)$$

where $\varepsilon$ is the magnitude of an average instantaneous energy gap of $H_S(t)$. Here $|g|$ measures an average coupling strength in (1.6), under the assumption that $S$ in (1.4) is chosen as dimensionless, of order 1, so that $|g|$ has units of energy. The cubic correction $O(\varepsilon^2|g|t^3)$ is present in the constant-$H_S$ case as well. The new commutator term that involves $H_g(t)$ is also of order $t^3$, provided $H_g(t)$ is well behaved at $t=0$, as can be verified by expanding it in powers of time. Thus, our short-time approximation for the overall evolution operator is

$$U_T(t) = W_S(t)e^{-i(H_B+H_{SB})t}W_S(t) , \qquad (2.9)$$

and the reduced density matrix of the system is given by

$$\rho_S(t) = Tr_B\left[ W_S(t)e^{-i(H_B+H_{SB})t}W_S(t)\rho(0)W_S^\dagger(t)e^{i(H_B+H_{SB})t}W_S^\dagger(t) \right]. \qquad (2.10)$$

In order to simplify this expression, we would like to factor out operators containing different bath modes. This will allow factorization of the trace to reduce it to single-mode traces. The bath mode operators enter in the exponents containing $H_{SB}$, as well as in the initial overall density matrix $\rho(0)$; see (1.4), (1.6) and (1.8). It proves convenient to replace the only system operator entering in these factors, namely, $S$ in the exponents, by c-numbers. This is done by introducing decomposition of the unit operator in terms of the eigenvectors of $S$, defined according to

$$S|\lambda\rangle = \lambda|\lambda\rangle . \qquad (2.11)$$

One can then derive the expression

$$\rho_S(t) = \sum_{\lambda\lambda'} e^{D_{\lambda\lambda'}(t)} W_S(t)|\lambda\rangle\langle\lambda|W_S(t)\rho_S(0)W_S^\dagger(t)|\lambda'\rangle\langle\lambda'|W_S^\dagger(t) , \qquad (2.12)$$

with

$$e^{D_{\lambda\lambda'}(t)} = Tr_B\left[ e^{-i(H_B+\lambda\sum_k X_k)t}\rho_B(0)e^{i(H_B+\lambda'\sum_k X_k)t} \right]. \qquad (2.13)$$

The trace in (2.13) can be factored into the product of traces over single bath modes. For a bath of bosonic modes, defined in (1.5) and (1.6), this calculation was already carried out in the literature,[25,33,73] with the result

$$D_{\lambda\lambda'}(t) = -(\lambda-\lambda')^2 \sum_k \frac{2|g_k|^2}{\omega_k}\sin^2(\omega_k t/2)\coth(\omega_k/2kT)$$
$$-i(\lambda^2-\lambda'^2)\sum_k \frac{|g_k|^2}{\omega_k}(\sin\omega_k t - \omega_k t) . \qquad (2.14)$$

We will refer to the powers in exponents of this type, entering the eigenstate sum in (2.12), as "decoherence functions," here $D_{\lambda\lambda'}(t)$. Note that the real parts of the decoherence functions are negative and, thus, correspond to decreasing magnitudes of the off-diagonal terms in the sum in (2.12). This effect drives the decoherence. The imaginary parts include phase distortion terms ($\sim \sin\omega t_k$), as well as terms linear in the time variable. In the time independent case these linear terms represent system energy gap changes induced by the bath.

Decoherence functions are usually evaluated in the continuum limit of infinitely many bath modes, by converting the summation over the modes to summation, and ultimately integration, over the mode frequencies. The "efficiency" of the modes of frequency $\omega$ in causing decoherence, is thus proportional not only to $|g(\omega)|^2$, but also to the density of states $\Upsilon(\omega)$. In model calculations,[60] one usually assumes a power-law behavior of the product $\Upsilon|g|^2$ at low frequencies, with a cut-off, $\omega_c$, at large frequencies, e.g.,

$$\sum_k |g_k|^2 f(\omega_k) \to \int_0^\infty (J\omega^n e^{-\omega/\omega_c}) f(\omega) d\omega \ . \tag{2.15}$$

## 3. Perturbative expansion for time-dependent gate interactions

Perturbative in $|g|$ expansions have been considered for time-independent interactions.[39,42,51-53] Many of them, however, have been unable to keep the unitarily of evolution resulting in unphysical probabilities at short times.[22] In this section, we introduce an approximation scheme that relies on a perturbative expansion in terms of the system-to-bath coupling strength $|g|$, while preserving unitarity. An advantage of having the alternative approximation accurate at short times is that by comparing the two approaches, one can gauge their limits of validity. A difficulty is that the resulting expressions are much more cumbersome[39,42] for analytical treatment than the short-time approach presented in the preceding section. Perturbative expansions in powers of $|g|$ are usually justified for quantum computing,[39,42,51-53] because only systems weakly coupled to the environment can achieve the desired threshold of $10^{-4}$, down to $10^{-6}$, error rate per cycle.[14,15]

The evolution of the overall density matrix $\rho(t)$ in the interaction representation is given by

$$i\dot\rho_I(t) = [S(t)\sum_k X_k(t), \rho_I(t)] \ , \tag{3.1}$$

where

$$\rho_I(t) \equiv U_S^\dagger(t) e^{iH_B t} \rho(t) e^{-iH_B t} U_S(t) \ , \tag{3.2}$$

$$S(t) \equiv U_S^\dagger(t) S U_S(t) \ , \tag{3.3}$$

and

$$X_k(t) \equiv e^{iH_B t} X_k e^{-iH_B t} \ . \tag{3.4}$$

An evolution operator can then be introduced via

$$i\dot U(t) = S(t)\sum_k X_k(t) U(t) \ , \tag{3.5}$$

such that

$$\rho_I(t) = U(t)\rho(0)U^\dagger(t) \ . \tag{3.6}$$

Note that $U(t)$ is directly related to $U_T(t)$ as suggested by (3.2), (3.6) and (2.2). Equation (3.5) can be used to develop approximations perturbative in the system-bath interaction. Furthermore, we will formulate such an approximation by utilizing a unitarity-preserving Magnus expansion.[64]

We note that it is always possible to write

$$U(t) = e^{-i\Omega(t)}, \tag{3.7}$$

where $\Omega(t)$ is Hermitian. Following the procedure of Ref. 65, one can expand $\Omega(t)$ in the power-series of the interaction strength, based on the fact that the right-hand side of (3.4) is linear in $|g|$; see (1.4) and (1.6). We have

$$\Omega(t) = \sum_{n=1}^{\infty} \Omega_n(t), \tag{3.8}$$

where the leading terms are

$$\Omega_1(t) = \int_0^t dt' \, S(t') \sum_k X_k(t'), \tag{3.9}$$

and

$$\Omega_2(t) = -\frac{i}{2} \int_0^t dt' \int_0^{t'} dt'' \, [S(t') \sum_{k'} X_{k'}(t'), S(t'') \sum_{k''} X_{k''}(t'')]. \tag{3.10}$$

Truncating expansion (3.8) at $n > 2$, we approximate evolution operator (3.7) to the second order in $|g|$.

Our ultimate goal will be to evaluate the trace of the overall density matrix over the bath modes,

$$\rho_S(t) = U_S(t) Tr_B \left[ e^{-iH_B t} U(t) \rho_S(0) \rho_B(0) U^\dagger(t) e^{iH_B t} \right] U_S^\dagger(t). \tag{3.11}$$

By writing the expression (3.10) in terms of the bosonic creation (annihilation) operators, we can identify terms that have no contribution to the trace in (3.11) within the second-order-in-$|g|$ approximation. For example, $g_k g_{k'} a_k^\dagger a_{k'}^\dagger$ and $g_{k''}^* g_{k'''}^* a_{k''} a_{k'''}$, originating from (3.10), should be paired up with terms with the appropriately matching indices in order to produce a nonzero contribution to the trace. Therefore, all such terms only contribute in order $|g|^4$ and can be omitted. One can also notice that terms $|g_k|^2 a_k^\dagger a_k$, when grouped with $\rho_B(0)$ under the trace, yield quantities that are small for low temperatures. The calculations of the trace of such combinations at low temperatures show that the resulting expressions are smaller than the leading order-$|g|^2$ contributions by factors of order $(kT/\omega_c)^n$, for the bath density of states (2.15). In our case $kT \ll \omega_c$, $n > 0$ (in most systems of interest even $n \geq 1$), and these terms are negligible.

After some rearrangement of terms in (3.9) and (3.10), we can replace, inside the trace only, the evolution operator (3.7) by the following expression, in which terms omitted in the exponent do not contribute to the trace in the leading orders of the approximations assumed,

$$U(t) \to e^{-i\sum_k F_k(t)X_k(t) - i\sum_k \tilde{F}_k(t)\tilde{X}_k(t) - i\sum_k G_k(t)|g_k|^2}, \tag{3.12}$$

where

$$F_k(t) \equiv \int_0^t dt' S(t')\cos\omega_k(t'-t), \tag{3.13}$$

$$\tilde{F}_k(t) \equiv \int_0^t dt' S(t')\sin\omega_k(t'-t), \tag{3.14}$$

$$\tilde{X}_k(t) = ig_k a_k^\dagger(t) - ig_k^* a_k(t), \tag{3.15}$$

$$G_k(t) \equiv -\frac{i}{2}\int_0^t dt' \int_0^{t'} dt'' [S(t'), S(t'')]\cos\omega_k(t'-t'')$$
$$-\frac{1}{2}\int_0^t dt' \int_0^{t'} dt'' \{S(t'), S(t'')\}\sin\omega_k(t'-t''). \tag{3.16}$$

The first two terms in the exponent of (3.12) came from (3.9); the third term originated from (3.10).

As in the short-time approximation case, the calculation of the trace over the bath modes can be carried out if we replace certain system operators by c-numbers. This can only be done if we first separate out the $k$-dependence of the system operators in (3.12). For a single-qubit system, which will be assumed from now on, we expand all the system operators introduced in (3.13), (3.14), (3.16), in terms of the Pauli matrixes $\sigma_j$, where $j \in \{x, y, z\}$,

$$F_k(t) = \sum_j f_k^j(t)\sigma_j, \quad \tilde{F}_k(t) = \sum_j \tilde{f}_k^j(t)\sigma_j, \quad G_k(t) \to \sum_j \gamma_k^j(t)\sigma_j. \tag{3.17}$$

When expanding $F_k(t)$ and $\tilde{F}_k(t)$, we did not include the identity (unit) operator, thus assuming, without loss of generality, that $Tr\, S = 0$. Indeed, the identity-operator term in $S$ does not affect the system-bath coupling and can be eliminated by redefining the bosonic bath mode coordinates.[26,57] The identity-operator term in $G_k(t)$ is irrelevant for the calculations of the density matrix to order $|g|^2$, as can be seen from (3.11) and (3.12). We note that for several-qubit systems, one should use an appropriate set of basic matrixes of higher dimension including identity. Terms containing different $\sigma_j$ in the exponent of (3.12) are next factorized by using the relation

$$\prod_j e^{gA_j} = e^{\sum_j gA_j + \frac{1}{2}g^2 \sum_{i<j}[A_i, A_j] + O(g^3)}, \tag{3.18}$$

where $A_j$ denote a set of operators, and $g$ is a small parameter. We get

$$U(t) = e^{-i\sigma_x \sum_k \left[ f_k^x X_k(t) + \tilde{f}_k^x \tilde{X}_k(t) + \gamma_k^x |g_k|^2 - |g_k|^2 \left( f_k^y f_k^z + \tilde{f}_k^y \tilde{f}_k^z \right) \right] + O(|g|^3)}$$
$$\times e^{-i\sigma_y \sum_k \left[ f_k^y X_k(t) + \tilde{f}_k^y \tilde{X}_k(t) + \gamma_k^y |g_k|^2 + |g_k|^2 \left( f_k^x f_k^z + \tilde{f}_k^x \tilde{f}_k^z \right) \right] + O(|g|^3)}$$
$$\times e^{-i\sigma_z \sum_k \left[ f_k^z X_k(t) + \tilde{f}_k^z \tilde{X}_k(t) + \gamma_k^z |g_k|^2 - |g_k|^2 \left( f_k^x f_k^y + \tilde{f}_k^x \tilde{f}_k^y \right) \right] + O(|g|^3)} \quad (3.19)$$

Finally, we insert the decomposition of unity in terms of the eigenvectors of the Pauli matrixes, i.e., $|x\rangle, |y\rangle$ and $|z\rangle$, with eigenvalues $x, y, z \in \{\pm 1\}$. These indexes, while allowing the use of the vector dot-product notation, see below, should no be confused with the Cartesian $x, y, z$ used in superscripts of the expansion coefficients in (3.17) and in labeling the Pauli matrixes. After some algebra utilizing properties of bosonic operators, we obtain the expression

$$U(t) = \sum_{\mathbf{x}} \mathfrak{M}_{\mathbf{x}} e^{-i\sum_k (\mathbf{x} \cdot \mathbf{f}_k) X_k(t)} e^{-i\sum_k (\mathbf{x} \cdot \tilde{\mathbf{f}}_k) \tilde{X}_k(t)}$$
$$\times e^{-i\sum_k |g_k|^2 \left[ \mathbf{x} \cdot (\mathbf{\gamma}_k - \mathbf{F}_k) - 2 \left( yx f_k^y \tilde{f}_k^x + zx f_k^z \tilde{f}_k^x + zy f_k^z \tilde{f}_k^y \right) \right] + O(g^3)}, \quad (3.20)$$

with $\mathfrak{M}_{\mathbf{x}} \equiv |x\rangle\langle x|y\rangle\langle y|z\rangle\langle z|$, $\mathbf{x} \equiv (x, y, z) = (\pm 1, \pm 1, \pm 1)$, where we defined

$$\mathbf{f}_k \equiv \left( f_k^x, f_k^y, f_k^z \right), \qquad \mathbf{\gamma}_k \equiv \left( \gamma_k^x, \gamma_k^y, \gamma_k^z \right), \quad (3.21)$$

and

$$\mathbf{F}_k \equiv \left( f_k^y f_k^z + \tilde{f}_k^y \tilde{f}_k^z, -f_k^x f_k^z - \tilde{f}_k^x \tilde{f}_k^z, f_k^x f_k^y + \tilde{f}_k^x \tilde{f}_k^y \right). \quad (3.22)$$

The evolution operator in (3.20) is still unitary, and the positivity of the density matrix is preserved. We point out that the use of the relation (3.18), with the operators $A_j$ proportional to the Pauli matrixes, makes the resulting expressions dependent on the choice of the Cartesian axis in the spin space, though the difference is of order higher than $|g|^2$.

The reduced density matrix is now easy to construct, by substituting (3.20) into (3.11),

$$\rho_S(t) = \sum_{\mathbf{x}, \mathbf{x}'} e^{D_{\mathbf{x}, \mathbf{x}'}(t)} U_S(t) \mathfrak{M}_{\mathbf{x}} \rho_S(0) \mathfrak{M}_{\mathbf{x}'}^\dagger U_S^\dagger(t). \quad (3.23)$$

Here the quantity

$$e^{D_{\mathbf{x}, \mathbf{x}'}(t)} = Tr_B \left[ e^{-i\sum_k (\mathbf{x} \cdot \mathbf{f}_k) X_k} e^{-i\sum_k (\mathbf{x} \cdot \tilde{\mathbf{f}}_k) \tilde{X}_k} \rho_B(0) e^{i\sum_k (\mathbf{x}' \cdot \tilde{\mathbf{f}}_k) \tilde{X}_k} e^{i\sum_k (\mathbf{x}' \cdot \mathbf{f}_k) X_k} \right]$$
$$\times e^{-i\sum_k |g_k|^2 \left[ (\mathbf{x} - \mathbf{x}') \cdot (\mathbf{\gamma}_k - \mathbf{F}_k) - 2 \left\{ (yx - y'x') f_k^y \tilde{f}_k^x + (zx - z'x') f_k^z \tilde{f}_k^x + (zy - z'y') f_k^z \tilde{f}_k^y \right\} \right]}, \quad (3.24)$$

introduces all the decoherence effects. Note that if we replace $D_{\mathbf{x}, \mathbf{x}'}(t) \to 0$, in (3.23), the resulting expression yields the coherent density matrix evolution obtained without coupling to the bath. The nonzero decoherence functions $D_{\mathbf{x}, \mathbf{x}'}(t)$ introduce "quantum noise" by perturbing both the magnitudes and phases of the terms in the sum in (3.23), which would otherwise combine to yield the coherent evolution.

The trace over the bath in (3.24) can be factored into a product of single-mode traces because the bath operators of different modes commute. The calculations are rather lengthy and involve coherent-state[26] and bosonic-operator-algebra techniques.[51] We only report the resulting expression,

$$D_{\mathbf{x},\mathbf{x}'}(t) = -\sum_k \frac{|g_k|^2}{2} \left[ \{(\mathbf{x}-\mathbf{x}')\cdot \mathbf{f}_k\}^2 + \{(\mathbf{x}-\mathbf{x}')\cdot \tilde{\mathbf{f}}_k\}^2 \right] \coth \frac{\omega_k}{2kT}$$
$$-i\sum_k |g_k|^2 \left[ (\mathbf{x}-\mathbf{x}')\cdot (\mathbf{\gamma}_k - \mathbf{F}_k) + \{(\mathbf{x}-\mathbf{x}')\cdot \mathbf{f}_k\}\{(\mathbf{x}+\mathbf{x}')\cdot \tilde{\mathbf{f}}_k\} \right] \quad (3.25)$$
$$+i\sum_k |g_k|^2 \left[ 2(yx-y'x') f_k^y \tilde{f}_k^x + 2(zx-z'x') f_k^z \tilde{f}_k^x + 2(zy-z'y') f_k^z \tilde{f}_k^y \right].$$

In the following sections we provide two examples of how the developed approximation schemes can be applied. Decoherence functions here are structured similarly to those in (2.14).

## 4. Exactly solvable model

In this section we discuss an adiabatic model, which is known to be exactly solvable.[26] For idling qubits, it is formulated as a model of system-bath interaction, such that no energy exchange is allowed between the system and environment. In other words, the Hamiltonian of the qubit system (1.2) is assumed to commute with the interaction part (1.4). Such models are also known as pure-decoherence and quantum-nondemolition processes.[66] For the time-dependent case, we could also consider models with $[H_S(t), S] = 0$. The operators involved are Hermitian and, when commuting, can be diagonalized in the same basis. For interactions involving operators $S$ that have no degenerate eigenstates, this assumption implies that $H_S(t)$ also commutes with itself at different times, and

$$U_S(t) = \exp\left[-i\int_0^t H_S(t')dt'\right]. \quad (4.1)$$

The eigenvectors of $H_S(t)$ and $S$ will be denoted as $|\lambda\rangle$, with the eigenvalues $E_\lambda(t)$ and $\lambda$, respectively.

The short-time approximation derived in Section II, is exact for the adiabatic model. However, the perturbative approach, developed in Section III, is approximate, and our explicit result (3.25) for its decoherence functions was derived for the single-qubit situation only. For a qubit—a two state system—a nontrivial interaction operator can be assumed proportional to some combination of Pauli matrixes, with symmetric values $\lambda \propto \pm 1$, and therefore its spectrum is indeed nondegenerate. Since all the Pauli matrixes are noncommuting, the only adiabatic model that can be constructed will have $H_S(t) = p(t)S$, where $p(t)$ is c-number. We also have $E_\lambda(t) = p(t)\lambda$.

The short-time approximation, which happens to be the exact solution, yields

$$\langle\lambda|\rho_S(t)|\lambda'\rangle = e^{-i(\lambda-\lambda')\int_0^t p(t')dt'}\langle\lambda|\rho_S(0)|\lambda'\rangle \\ \times e^{-(\lambda-\lambda')^2\sum_k \frac{2|g_k|^2}{\omega_k}\sin^2(\omega_k t/2)\coth(\omega_k/2kT)}. \quad (4.2)$$

In the adiabatic model, the diagonal density matrix elements remain unchanged. The process of decoherence corresponds to decay of the off-diagonal matrix elements, since the sum over $k$ in the second exponent is a monotonically increasing function of time.

To calculate the reduced density matrix using the perturbative approach we notice that $G_k(t)$ in (3.16) vanishes, while (3.13) and (3.14) give

$$\mathbf{f}_k(t) = \frac{1}{2}Tr(\boldsymbol{\sigma}S)\frac{\sin\omega_k t}{\omega_k} \quad, \quad \tilde{\mathbf{f}}_k(t) = \frac{1}{2}Tr(\boldsymbol{\sigma}S)\frac{\cos\omega_k t - 1}{\omega_k}, \quad (4.3)$$

where $\boldsymbol{\sigma}$ is the vector of the Pauli matrixes. The reduced density matrix is

$$\langle\lambda|\rho_S(t)|\lambda'\rangle = e^{-i(\lambda-\lambda')\int_0^t p(t')dt'}\sum_{\mathbf{x},\mathbf{x}'}\langle\lambda|\mathfrak{M}_\mathbf{x}\rho_S(0)\mathfrak{M}_{\mathbf{x}'}^\dagger|\lambda'\rangle e^{D_{\mathbf{x},\mathbf{x}'}(t)}, \quad (4.4)$$

where $\mathfrak{M}_\mathbf{x}$ was introduced in connection with (3.20) and can be written as

$$\mathfrak{M}_\mathbf{x} = \frac{(1+ixy)[1+z-iy(1-z)]}{16}\begin{pmatrix} 1+z & 1-z \\ x(1+z) & x(1-z) \end{pmatrix}; \quad (4.5)$$

we recall that here $x, y, z \in \{\pm 1\}$.

The decoherence functions $D_{\mathbf{x},\mathbf{x}'}(t)$ are defined by (3.25). The real part is

$$\mathrm{Re}\, D_{\mathbf{x},\mathbf{x}'}(t) = -\sum_k \frac{2|g_k|^2}{\omega_k^2}\left[\frac{1}{2}Tr(\boldsymbol{\sigma}S)\cdot(\mathbf{x}-\mathbf{x}')\right]^2 \sin^2\frac{\omega_k t}{2}\coth\frac{\omega_k}{2kT}. \quad (4.6)$$

The expression for the imaginary part, also proportional to $|g|^2$, is lengthy and is not reproduced here. The property $\sum_\mathbf{x}\mathfrak{M}_\mathbf{x} = 1$ guarantees that for $|g|=0$, the coherent dynamics is reproduced in (4.4).

As mentioned earlier, the present approximation depends, in orders higher that $|g|^2$, on the choice of the axis in the spin space. This property is illustrated in (4.4) as follows. One can show that for $S = \sigma_x, \sigma_y, \sigma_z$, the imaginary part of $D_{\mathbf{x},\mathbf{x}'}(t)$ is zero, and the real part also simplifies so that the summation in (4.4) reproduces the exact solution (4.2) for the adiabatic model. For a general choice of $S$, the perturbative approach produces only approximate results with the expected order-$|g|^3$ deviation, as will be illustrated in the following calculations.

It is convenient to compare (4.2) and (4.4) by evaluating the measure of decoherence introduced in Ref. 32. It is formulated as a norm of the deviation of the reduced density matrix from the coherent density matrix, $\rho_C(t)$, of the system,

$$\chi(t) = \rho_S(t) - \rho_C(t). \quad (4.7)$$

The measure is defined in terms of the eigenvalues $\varsigma_j(t)$ of $\chi(t)$,

$$\|\chi(t)\|_\lambda = \max_j |\varsigma_j(t)| \ . \tag{4.8}$$

In Ref. 32, the measure, $\|\chi(t)\|_\lambda$, was further maximized over all the possible initial states of the system, i.e., over $\rho_S(0)$, in order to suppress oscillations at the system's own energy-gap frequencies. For the purposes of comparing the two approximation schemes, a simpler definition (4.8) is sufficient.

Our illustrative calculations were carried out for the measure (4.8), assuming the Ohmic bath, i.e., $n = 1$ in (2.15), and for $T = 0$. The latter choice allows analytical evaluation of the integrals. Moreover, for short and intermediate times, small non-zero temperature and $T = 0$ results are not too different. We also use spherical coordinates[52] to define $S$ and $\rho_S(0)$ in the qubit space,

$$S = \cos\varphi_s \sin\theta_s \sigma_x + \sin\varphi_s \sin\theta_s \sigma_y + \cos\theta_s \sigma_z \ , \tag{4.9}$$

and

$$\rho_S(0) = \frac{1}{2}\left(1 + \cos\varphi_\rho \sin\theta_\rho \sigma_x + \sin\varphi_\rho \sin\theta_\rho \sigma_y + \cos\theta_\rho \sigma_z\right) \ . \tag{4.10}$$

Figure 1 shows $\|\chi(t)\|_\lambda$ for the adiabatic model, calculated using (4.2) and (4.4). Since for the adiabatic model the short-time approach gives the exact solution, the deviation, also shown in Fig. 1, illustrates the error in the perturbative approximation. We also give illustrative values for different bath density of states in Table 1, where $\|\chi(t)\|_\lambda$ for a certain $t$ value was calculated for Ohmic, $n = 1$, and two super-Ohmic, $n = 2, 3$, cases, see (2.15).

## 5. Decoherence of a two-state system in a rotating wave

In this section we consider an example of time-dependent gate control, introduced via external rotating wave.[66] This model is well known and found broad applications.[66,67] Without an environmental noise, the model allows an exact solution and the evolution operator can be obtained explicitly. We set the Hamiltonian of an idling qubit to

$$H_i = a\sigma_z \ . \tag{5.1}$$

Here $a$ is half the energy gap between the two qubit levels. The gate control is represented by

$$H_g(t) = c\left(\sigma_x \cos 2at + \sigma_y \sin 2at\right) \ , \tag{5.2}$$

where $c$ is the amplitude of the external wave. The wave frequency is tuned in resonance with the qubit. The coherent system evolution operator for (5.1) and (5.2) can be derived as

$$U_S(t) = e^{-iat\sigma_z} e^{-ict\sigma_x} \ . \tag{5.3}$$

Such a gate control can be used, for example, to represent phase and NOT gate functions[46] applied consecutively. Indeed, applying $U_S(t)$ in the form (5.3) for time $t = \pi/2a$ and $c = a$, one switches the state $A|+\rangle + B|-\rangle$ to $A|-\rangle - B|+\rangle$, where $|+\rangle, |-\rangle$ are the eigenstates of the idling two-state system with the Hamiltonian (5.1).

For simplicity, we set the system-bath interaction term to be commuting with $H_i$, to represent, if the qubit were idling, the so-called "pure phase noise",

$$S = \sigma_z \ . \tag{5.4}$$

In the interaction representation, we have

$$S(t) = \sigma_z \cos 2at + \sigma_y \sin 2at \ . \tag{5.5}$$

We begin with the short time approximation. For the bath density of states given by (2.15), the integrals in the expression for the decoherence functions $D_{\lambda\lambda'}(t)$ for $T = 0$ can be calculated analytically. For the Ohmic case, $n = 1$, we get

$$D_{\lambda\lambda'}(t) = -\frac{1}{2} J (\lambda - \lambda')^2 \ln\left(1 + \omega_c^2 t^2\right) \ . \tag{5.6}$$

For $T > 0$, the decoherence functions have been described in the literature; their general properties are known,[25,33,60,73] but no closed-form expressions are available. The explicit expression for $W_S(t)$ can be obtained, but it is cumbersome and is not shown here.

Expressions for the decoherence functions for the perturbative approximation are not available analytically even at zero temperatures due to the complexity of $\mathbf{f}_k$, $\tilde{\mathbf{f}}_k$ and $\gamma_k$. These functions, however, do not have any singularities or branch cuts, making $D_{\mathbf{xx'}}(t)$ easy to compute numerically.

Let us begin by considering the effect of quantum noise on what would be the gate-induced oscillation between the $|+\rangle$ and $|-\rangle$ states of an idling qubit in the course of the combined "phase and NOT" gate function described following (5.3). In Fig. 2, we plot the deviation of the actual probability for the qubit to be in the state $|+\rangle$, i.e. $\rho_{S,++}(t) \equiv \langle +|\rho_S(t)|+\rangle$, from the coherent-evolution value, i.e., $\rho_{C,++}(t) \equiv \langle +|\rho_C(t)|+\rangle$. The probability of the $|+\rangle$ state, $\rho_{C,++}(t)$, is also shown. Fig. 2 demonstrates that up to the time of approximately one idling qubit oscillation, the two approximations yield virtually indistinguishable results. Eventually, however, they diverge from each other significantly. This observation supports the analysis of the omitted corrections made for the short-time approach, see (2.8), and the discussion in Section II. Indeed, in the case of (5.1) and (5.2), $\varepsilon$ in (2.8) is of order $a, c$, which are 1 in dimensionless units of Fig. 2.

Note that Fig. 2 does not show the phase changes (the off-diagonal matrix elements) which are also affected by the decoherence processes. The overall density matrix deviates from coherent even when the deviation for the probability happens to vanish, e.g., for the perturbative approximation at a time close to $1.8/a$ in Fig. 2. To show this, we plot the measure of decoherence $\|\chi(t)\|_\lambda$, see (4.8), in Fig. 3, for the same set of parameters. The short time approximation is consistent with the perturbative one up to the times of order $1/a$ and $1/c$, as predicted in Section II, see (2.8).

It is instructive to examine the structure of the decoherence functions of the perturbative approximation $D_{\mathbf{xx'}}(t)$. There are total 64 functions $D_{\mathbf{xx'}}(t) = D_{x=\pm 1, y=\pm 1, z=\pm 1; x'=\pm 1, y'=\pm 1, z'=\pm 1}(t)$. Many of them coincide, and some are zero.

The nonzero ones are presented on Fig. 4 for the same parameters as used for Fig. 2, and a somewhat longer time interval. The real parts of the nonzero $D_{\mathbf{xx'}}(t)$ separate into two bands, as illustrated in Fig. 4, growing on average linearly with time, each with superimposed oscillations. This defines two different time scales for the decoherence processes, see (4.7) and (3.23). The reason of the separation may be understood if one investigates the structure of the first sum in (3.25). When we deal with the adiabatic model, $[U_S(t), S] = 0$, $\mathbf{f}_k$ and $\tilde{\mathbf{f}}_k$ differ only in magnitude, which results in factorization of $\{(\mathbf{x} - \mathbf{x'}) \cdot Tr(\mathbf{\sigma} S)\}^2$ before the integration over the bath modes. This leads to decoherence functions with relative values of the real parts determined by the form of $S$. For example when $S = \sigma_z$, one obtains a single non-zero value. Noncommutativity of $U_S(t)$ and $S$ introduces a qualitatively different separation. It is now caused by the difference in integrating the two terms of the first sum in (3.25) over the bath modes. In the present example we see, that initially when the short-time and perturbative approaches agree (Fig. 3), the two bands (Fig. 4) has no clear separation. The imaginary part shows a set of linearly growing phases with superimposed small oscillations, as expected from the structure of (3.25). The contribution of each band of $\operatorname{Re} D_{\mathbf{xx'}}(t)$ to decoherence depends on the choice of the initial density matrix, with which they are interconnected through $\mathfrak{M}_{\mathbf{x}}$ and $\operatorname{Im} D_{\mathbf{xx'}}(t)$, see (3.23).

Next, let us analyze a "strong field" scenario, by setting the amplitude of the rotating wave to, e.g., $c = 15a$. The measure of decoherence calculated within our two approximations is shown in Fig. 5, for two different time scales. Since the general expectation is that the effects of quantum noise should increase with time, the results of the short-time approximation, curve (a) in Fig. 5, suggest that it breaks down here earlier than in the case $c = a$. This breakdown occurs for times of order $1/c$, for which the short-time approximation also visibly deviates from the perturbative one. For larger times, the short-time approach clearly underestimates decoherence. Heuristic observations based on the monotonic behavior, Fig. 5, and the adiabatic-case calculations of the preceding section, suggest that the perturbative approximation has a larger range of validity in this strong-gate case.

More detailed analysis of the calculations with different values of the rotating wave amplitude is presented on Fig. 6, where we show the decoherence functions of the perturbative approach for varying $c$. Both bands of the nonzero $\operatorname{Re} D_{\mathbf{xx'}}(t)$ values increase with increasing $c$, until $c$ reaches approximately half of $\omega_c$. At that point they begin decreasing, as do the imaginary parts of the decoherence functions. Physically, such a phenomenon can be explained as follows. As the system dynamics, driven by the external wave, becomes faster than the cutoff frequency of the bath modes, $\omega_c$, the resonant, energy-exchange relaxation processes are suppressed. The pure-decoherence type, phase noise still remains, and is captured correctly by the perturbative approximation, whereas we point out that the short-time approximation is not designed to describe this regime.

In summary, we developed two approximation schemes for evaluating decoherence for gate-controlled quantum systems subject to quantum noise described by a bath of bosonic modes. The short-time approximation is simpler and more versatile, but has a somewhat more limited range of validity, than the perturbative approximation. The latter approximation was worked out in detail only for a single-qubit system. Illustrative calculations were reported for both approximations for the case of the exactly solvable adiabatic model and for the rotating-wave quantum-gate model.

**Acknowledgments**

The authors acknowledge support of this research by the National Security Agency and Advanced Research and Development Activity under Army Research Office contract DAAD-19-02-1-0035, and by the National Science Foundation, grant DMR-0121146.

**TABLES**

|       | short-time approximation | perturbative approximation | the deviation |
|-------|--------------------------|----------------------------|---------------|
| $n = 1$ | $3.48431 \times 10^{-6}$ | $3.48429 \times 10^{-6}$ | $2.41771 \times 10^{-11}$ |
| $n = 2$ | $3.06940 \times 10^{-5}$ | $3.06921 \times 10^{-5}$ | $1.87601 \times 10^{-9}$ |
| $n = 3$ | $9.22894 \times 10^{-4}$ | $9.21203 \times 10^{-4}$ | $1.69110 \times 10^{-6}$ |

**Table 1.** Measure of decoherence for the adiabatic model, for the short-time and perturbative approaches, calculated for the Ohmic ($n=1$) and super-Ohmic ($n=2,3$) baths. The deviation shows the expected third-order behavior. The values shown are for $t=1$, for the same set of parameters and choice of units as in Fig. 1.

**FIGURES**

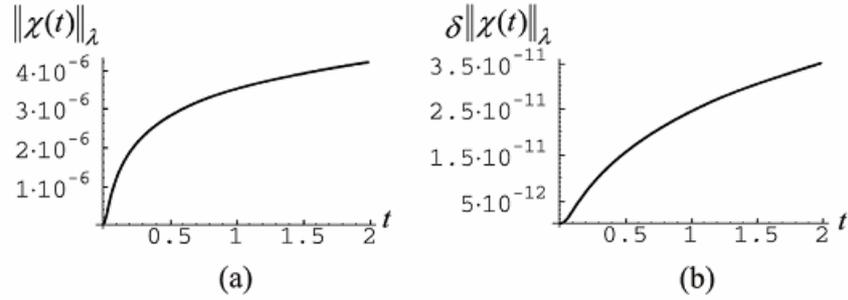

**Figure 1.** (a) Measure of decoherence, $\|\chi(t)\|_\lambda$, for the short-time approach in the adiabatic case (exact solution), and (b) the deviation, $\delta\|\chi(t)\|_\lambda$, of measure of decoherence calculated with the perturbative approximation from the one computed for the exact solution. The magnitude of the deviation is consistent with order $|g|^3 \sim J^{3/2}$ or better. The parameter values used were $J = 10^{-6}$, $\omega_c = 30$, $T = 0$, $\varphi_s = 0.16\pi$, $\theta_s = 0.35\pi$, $\varphi_\rho = 1.33\pi$, $\theta_\rho = 0.75\pi$. The time and energies are given in terms of half the idling qubit energy gap.

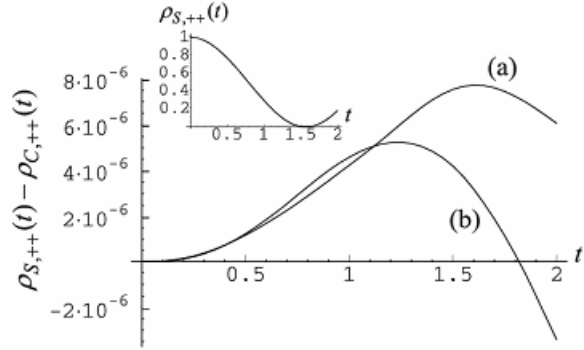

**Figure 2.** The deviation of the probability for the system to remain in the $|+\rangle$ state for noisy, $\rho_{S,++}(t)$, vs. coherent, $\rho_{C,++}(t)$, dynamics, for a rotating wave gate function. Curve (a) represents the short-time approximation, while curve (b) represents the perturbative approximation. The system is initially in the $|+\rangle$ state, and the parameters are as follows, $J = 10^{-6}$, $a = c = 1$, $\omega_c = 30$, $T = 0$. Time and energies are given in terms of half the idling qubit energy gap. For convenience we also show $\rho_{S,++}(t)$ on the same time scale (the curves for the two approximations are indistinguishable for the times shown).

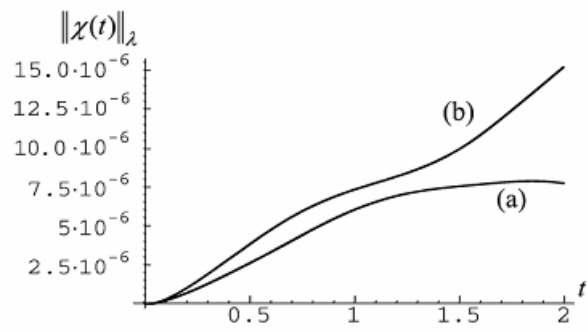

**Figure 3.** Measure of decoherence calculated with the (a) short-time, and (b) perturbative approximations for the rotating wave gate. The system and all the parameters are the same as in Fig. 2.

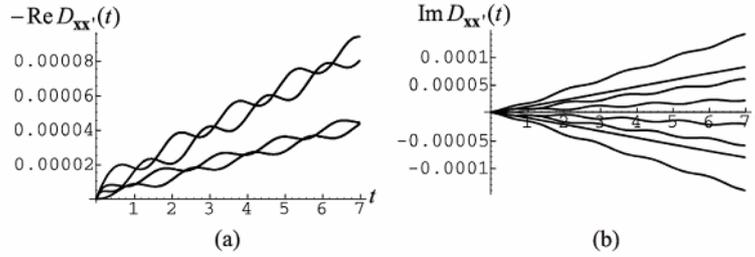

**Figure 4.** Nonzero decoherence functions, $D_{x,x'}(t)$, of the perturbative approximation for the rotating wave gate model. The formation of two distinct bands in the real part of $D_{x,x'}(t)$ is observed. The imaginary parts are linear in time with small oscillations superimposed. The parameters are as follows, $J = 10^{-6}$, $a = c = 1$, $\omega_c = 30$, $T = 0$. Time and energies are given in terms of half the idling qubit energy gap.

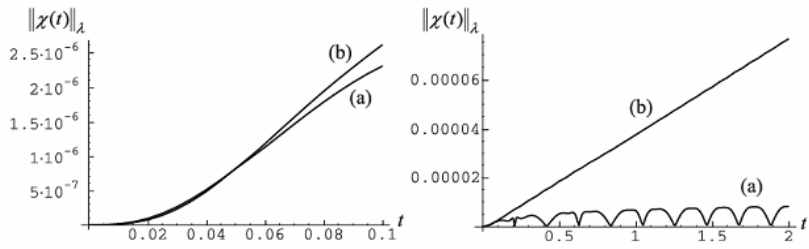

**Figure 5.** Measure of decoherence calculated using the (a) short-time, and (b) perturbative approximations for the rotating wave gate with a large amplitude. Two time scales are shown. The system is initially in the $|+\rangle$ state and the parameters are as follows, $J = 10^{-6}$, $a = 1$, $c = 15$, $\omega_c = 30$, $T = 0$. Time and energies are given in terms of half the idling qubit energy gap.

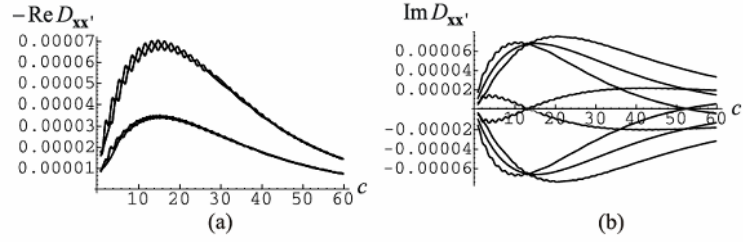

**Figure 6.** The (a) real and (b) imaginary parts of decoherence functions, $D_{x,x'}(t)$, of the perturbative approximation at $t=1$, for varying values of the rotating wave amplitude, $c$. The parameters are as follows, $J=10^{-6}$, $a=1$, $\omega_c=30$, $T=0$. Time and energies are given in terms of half the idling qubit energy gap.